\documentclass[twocolumn]{aastex631}

\usepackage{multirow}
\usepackage{float}

\shorttitle{Detectability of CO\textsubscript{2}-CH\textsubscript{4} on TRAPPIST-1e}
\shortauthors{Rotman et al.}

\graphicspath{{./}{figures/}}

\begin{document}

\title{GCM Constraints on the Detectability of the CO\textsubscript{2}-CH\textsubscript{4} Biosignature Pair on TRAPPIST-1e with JWST}

\author[0000-0003-4459-9054]{Yoav Rotman}
\affiliation{School of Earth and Space Exploration, Arizona State University, 875 S. Palm Walk, Tempe, AZ 85281, USA}
\affiliation{Department of Astronomy, University of Maryland, 4296 Stadium Drive, College Park, MD 20742, USA}

\author[0000-0002-9258-5311]{Thaddeus D. Komacek}
\affiliation{Department of Astronomy, University of Maryland, 4296 Stadium Drive, College Park, MD 20742, USA}

\author[0000-0002-2662-5776]{Geronimo L. Villanueva}
\affiliation{NASA Goddard Space Flight Center, 8800 Greenbelt Road, Greenbelt, MD 20771, USA}

\author[0000-0002-5967-9631]{Thomas J. Fauchez}
\affiliation{NASA Goddard Space Flight Center, 8800 Greenbelt Road, Greenbelt, MD 20771, USA}

\author[0000-0002-2739-1465]{Erin M. May}
\affiliation{Johns Hopkins APL, 11100 Johns Hopkins Rd, Laurel, MD 20723, USA}



\begin{abstract}

Terrestrial exoplanets such as TRAPPIST-1e will be observed in a new capacity with JWST/NIRSpec, which is expected to be able to detect 
CO\textsubscript{2}, CH\textsubscript{4}, and O\textsubscript{2} signals, if present, 
with multiple co-added transit observations. The CO\textsubscript{2}-CH\textsubscript{4} pair in particular is theorized to be a potential biosignature when inferred to be in chemical disequilibrium. Here, we simulate TRAPPIST-1e's atmosphere using the ExoCAM General Circulation Model (GCM), assuming an optimistic haze-free, tidally locked planet with an aquaplanet surface, with varying atmospheric compositions from $10^{-4}$ bar to 1 bar of partial CO\textsubscript{2} pressure with 1 bar of background N\textsubscript{2}. 
We investigate cases both with and without a modern Earth-like CH\textsubscript{4} mixing ratio to examine the effect of CO$_2$ and CH$_4$ on the transmission spectrum and climate state of the planet.
We demonstrate that in the optimistic haze-free cloudy case, H\textsubscript{2}O, CO\textsubscript{2}, and CH\textsubscript{4} could all be detectable in less than 50 transits within an atmosphere of 1 bar N\textsubscript{2} and 10 mbar CO\textsubscript{2} during JWST’s lifespan 
with NIRSpec as long as the noise floor is $\lesssim$ 10 ppm. We find that in these optimistic cases, JWST may be able to detect potential biosignature pairs such as CO\textsubscript{2}-CH\textsubscript{4} in TRAPPIST-1e's atmosphere across a variety of atmospheric CO$_2$ content, and that temporal climate variability does not significantly affect spectral feature variability for NIRSpec PRISM. 

\begin{center}
(Accepted to ApJL, Dec. 9, 2022)
\end{center}
\end{abstract}



\section{Introduction} \label{sec:intro}

JWST is expected to have the capability to characterize the atmospheric composition of temperate rocky planets. In particular, the NIRSpec PRISM instrument
will likely have the ability to identify spectral features in the atmospheres of temperate terrestrial exoplanets in 
as few as ten transits \citep{batalha_transiting_2014, batalha_strategies_2018, fauchez_impact_2019, birkmann_near-infrared_2022}. NIRSpec's broad wavelength range of $0.6-5.3$ $\mu$m
and moderate resolution of $R \approx 20-300$ \citep{jakobsen_near-infrared_2022} make it ideal for constraining spectral features of multiple habitability indicators and biosignatures including H$_2$O, CO$_2$, and CH$_4$
\citep{fauchez_impact_2019}. Simultaneous measurements with NIRSpec PRISM across a broad wavelength range, which would include multiple spectral features at different wavelengths, 
can enhance the signal-to-noise ratio of these species. 

TRAPPIST-1 is an ultracool M8V dwarf star
($T_{eff} = 2550 \pm 50 $ K) at a distance of twelve parsecs, with seven known planetary companions \citep{liebert_ri_2006, gillon_temperate_2016, gillon_seven_2017}. Of these, planets e, f, and g are all potentially within the habitable zone of their planet, although f and g would require atmospheres with a greenhouse gas complement beyond H$_2$O, e.g., with CO$_2$ and/or CH$_4$, due to their lower received flux \citep{kasting_habitable_1993, kopparapu_habitable_2013, gillon_seven_2017, lincowski_evolved_2018}. However, e has previously been shown to feasibly sustain some amount of liquid surface water under a much wider variety of atmospheric surface pressures than f or g \citep{turbet_modeling_2018, sergeev_trappist-1_2022}, and is among the most likely known exoplanets to host surface liquid water \citep{wolf_assessing_2017}. It will be observed throughout four transits with NIRSpec PRISM during JWST's Cycle 1 (GTO 1331, PI: Lewis), which may be enough to partially characterize its atmosphere \citep{morley_observing_2017, fauchez_impact_2019,  lustig-yaeger_detectability_2019}, and may in the future be complemented by reflection and emission spectra from ELT \citep{lin_high-resolution_2022}. JWST is expected to be able to view TRAPPIST-1e transiting a maximum of 85 times during its nominal 5.5 year lifetime \citep{lustig-yaeger_detectability_2019, komacek_clouds_2020}. However, with the recent increase of JWST's potential lifetime 
to 20 years \citep{rigby_characterization_2022}, TRAPPIST-1e may be viewable up to $\sim$320 times during transit, well past the required number of transits predicted
to identify molecular spectral signatures \citep{lustig-yaeger_detectability_2019, komacek_clouds_2020}.

TRAPPIST-1e has a near-Earth like radius and density, and receives 60\% of Earth's incident flux \citep{gillon_seven_2017}. It is likely either tidally locked or in a higher-order orbital resonance \citep{turbet_modeling_2018}, with an orbital period of $6.099$ days \citep{gillon_seven_2017}.
TRAPPIST-1e may be an aquaplanet, as shown by \cite{agol_refining_2021}, which identified TRAPPIST-1e's density as $\rho = 0.889^{+0.030}_{-0.033}\rho_{E}$ using transit-timing variations, and found that if iron makes up $\ge$25\% of the total planetary mass, there should be a higher water mass fraction than Earth's to account for the calculated density. 
Observations to date have only ruled out a hydrogen-rich atmosphere for TRAPPIST-1e,
with a lack of strong spectral features observed with HST/WFC3
implying a terrestrial planet with a higher mean molecular weight atmosphere \citep{de_wit_atmospheric_2018}. If TRAPPIST-1e is tidally locked and sustains surface liquid water, water vapor signal strength in transit spectroscopy may be dependent on cloud cover, as
water clouds would flatten the spectral features by increasing the continuum height, requiring more transits to identify spectral features with certainty \citep{greene_characterizing_2016, fauchez_impact_2019, suissa_dim_2020, mikal-evans_detecting_2021}. 

It has been previously demonstrated that, similar to the O\textsubscript{2}$-$CH\textsubscript{4} biosignature pair \citep{krissansen-totton_disequilibrium_2018, meadows_exoplanet_2018}, CO\textsubscript{2}$-$CH\textsubscript{4} could serve as a 
biosignature when seen in disequilibrium in a planetary atmosphere \citep{kleinbohl_buildup_2018, mikal-evans_detecting_2021}. Life could also drive the atmospheric chemistry to equilibrium via catalysis. However, in order to study the potential to observationally assess the inhabitance of a planet with JWST, we focus here on the previously proposed CO$_2$-CH$_4$ disequilibrium pair as a possible biosignature \citep{meadows_exoplanet_2018}. The O\textsubscript{2}$-$CH\textsubscript{4} pair is harder to characterize as a biotic reaction for TRAPPIST-1e due to the likelihood of O\textsubscript{2}/O\textsubscript{3} false positives in transit spectra from pre-main sequence runaway greenhouse effects in water-rich planetary atmospheres orbiting M dwarfs \citep{tian_high_2014,wordsworth_abiotic_2014,luger_extreme_2015}, making the CO\textsubscript{2}$-$CH\textsubscript{4} pair a better candidate to constrain the potential biosphere of TRAPPIST-1e.
A lack of CO in a CH\textsubscript{4}-inclusive environment is only known to exist if that methane is produced biotically, as other methane sources (such as volcanic outgassing and high-frequency impact events) contain carbon monoxide as a byproduct \citep{krissansen-totton_understanding_2022}. As such, atmospheres with significant CO\textsubscript{2} and CH\textsubscript{4} but little CO 
have the potential to be maintained biotically.


Previous work has shown that climatological factors such as the cloud cover and day-night heat transport on tidally locked rocky exoplanets depend strongly on the coupled planetary parameters of rotation rate and instellation 
 \citep{yang_stabilizing_2013,wolf_assessing_2017, turbet_modeling_2018, fauchez_impact_2019, shields_climates_2019, suissa_dim_2020}. For tidally locked planets such as TRAPPIST-1e, it is expected that the climate is characterized by day-night temperature contrasts, with a hot and cloudy substellar point and a 
 cold and less cloudy antistellar point \citep{merlis_atmospheric_2010,showman_atmospheric_2013-1, yang_stabilizing_2013, koll_temperature_2016}. TRAPPIST-1e in particular is a special case, that falls between the Rhines (or intermediate) rotator and the fast rotator regimes
 \citep{showman_atmospheric_2013-1, haqq-misra_demarcating_2018, sergeev_bistability_2022}. As such, it is expected to contain aspects of both regimes, depending on atmospheric composition; while the day-night temperature contrast should be large as in intermediate-rotating tidally-locked planets, there may also be a superrotating 
 eastward jet in the tropics, 
 with large cyclonic eddies in the extratropics formed by the day-night contrast causing a global standing wave pattern, as expected on a fast rotator \citep{showman_atmospheric_2013-1}. In this case, the weak temperature gradient (WTG) parameter, which encapsulates the effect of rotation on planetary-scale atmospheric dynamics, is $\Lambda \approx 2$, as predicted in  \citet{pierrehumbert_atmospheric_2019} for a Rhines rotator tidally-locked planet $\left( \Lambda > 1 \right)$. However, the climate state of this model includes a superrotating equatorial jet, as predicted for a fast rotator, which arises from TRAPPIST-1e's unique bistablity given its location at the edge of the fast and Rhines rotator regimes \citep{haqq-misra_demarcating_2018, sergeev_bistability_2022}.

Significant temporal variability is expected in the climates of tidally-locked planets with rotation periods similar to TRAPPIST-1e, although its mechanism may be difficult to identify, and is likely related to planetary-scale wave propagation \citep{pierrehumbert_atmospheric_2019, song_asymmetry_2021, cohen_traveling_2022}. Additionally, a longitudinally-asymmetric stratospheric oscillation (LASO), analogous to Earth's quasi-biennial oscillation (QBO) and caused by vertical propagation of gravity waves,
emerges in GCMs of Proxima Centauri b, an intermediate-rotating tidally locked rocky planet \citep{cohen_longitudinally_2021}. 
As such, there may be a
temporal variation in the stratospheric temperature and humidity the resulting cloud cover, which could amplify the expected variability in transmission signal strength at the limb. However, any east-west limb asymmetry is likely small and undetectable with JWST \citep{song_asymmetry_2021}.

For tidally locked planets exhibiting temporal variability, it is key to study the three-dimensional circulation; water vapor, cloud formation, and horizontal heat transport all require a three-dimensional model to properly simulate \citep{Joshi:1997,pierrehumbert_palette_2011,fauchez_impact_2019,may_water_2021}. Likewise, for simulating transmission spectra of tidally locked planets, limb conditions are particularly important, which can only be consistently predicted along with dayside and nightside properties using a 3D GCM. Rotation rate of the planet also dictates the circulation regime \citep{Noda:2017aa, krissansen-totton_detectability_2018} and cloud transport \citep{showman_atmospheric_2013-1}, which cannot be sufficiently modeled in one dimension. As such, our motivation in this work requires use of a GCM rather than a one-dimensional model.

This work follows this structure: in Section \ref{sec:method}, we discuss our GCM simulations and parameters used, as well as the post-processing through the NASA Planetary Spectrum Generator and subsequent analysis. In Section \ref{sec:results}, we discuss transmission spectrum results for the 10\textsuperscript{-2} bar pCO\textsubscript{2} case, their dependence on atmospheric climatology, and overarching trends across our full set of atmospheric cases with varying pCO$_2$ and including/excluding CH$_4$. In Section \ref{sec:discussion}, we discuss the implications of our findings on biosignature detectability with JWST and their limitations. Finally, in Section \ref{sec:conclusion}, we summarize our results and their significance for future transmission spectroscopy of TRAPPIST-1e with JWST. 

\begin{figure*}[]
    \centering
    \hspace*{-0.33cm}
    \includegraphics[width=185mm]{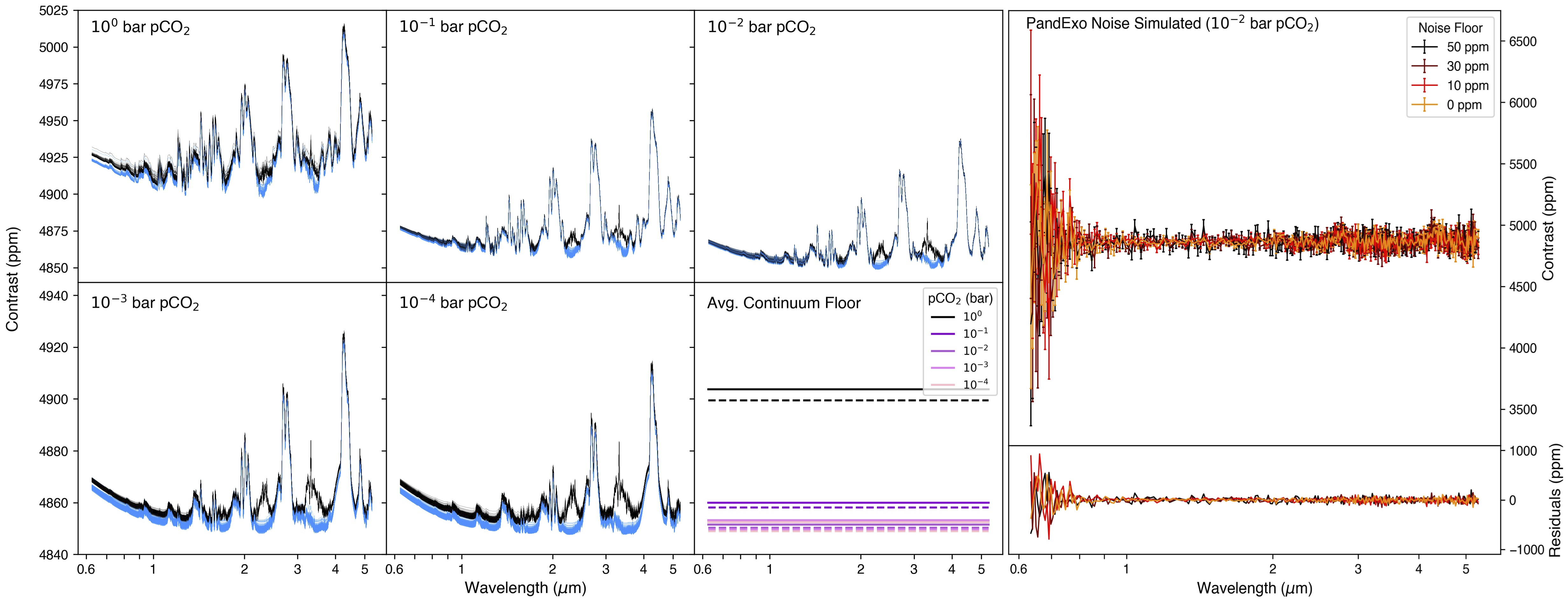}
    \caption{Transmission spectra for a full year of the 1, $10^{-1}$, $10^{-2}$, $10^{-3}$, and $10^{-4}$ bar pCO\textsubscript{2} cases (left), and a resulting PandExo noise output example (right). Black lines represent the methane case, while blue lines represent the methane-free case for each pCO\textsubscript{2}. Higher carbon dioxide cases tend to have larger features but also a higher continuum floor, which masks smaller features like those of CH\textsubscript{4}, as seen in the 1 bar case; note that the $10^{-3}$ and $10^{-4}$ bar cases are shown on a smaller vertical scale to emphasize spectral features. The bottom right panel shows the time-averaged continuum floor of all 360 spectra with (solid line) and without (dashed line) methane, for each of the five pCO\textsubscript{2} cases. The height of continuum floors and the difference between the methane-inclusive and exclusive continuum floors for each case both increase dramatically with increasing pCO\textsubscript{2}. The four varied noise floors (0 through 50 ppm) are shown in the right plot for a standard $10^{-2}$ bar pCO$_2$ spectrum, with the residuals plotted below. Higher noise floors correspond to larger residuals, and the sub-$1$ $\mu$m region has larger uncertainties than the $1-5$ $\mu$m region for all noise floors.}
    \label{fig:spectratotal}
\end{figure*}

\section{Methods} \label{sec:method}
\subsection{Atmospheric Model} \label{sec:model}
Building upon previous modeling work by \cite{may_water_2021}, we simulate the climate dynamics of TRAPPIST-1e using the \verb|ExoCAM| GCM\footnote{\url{https://github.com/storyofthewolf/ExoCAM}} \citep{wolf_exocam_2022}. \verb|ExoCAM| utilizes the \verb|ExoRT|\footnote{\url{https://github.com/storyofthewolf/ExoRT}} correlated-k radiative transfer scheme, which uses the \verb|HITRAN| absorption database and the \verb|HELIOS-K| opacity calculator \citep{rothman_hitran_2005, grimm_helios-k_2015, malik_helios_2017, grimm_helios-k_2021}.
We use a $4^{\circ} \times 5^{\circ}$ horizontal GCM resolution with 40 atmospheric layers and 28 correlated-k bins in the radiative transfer model. 
We simulate a grid of atmospheric compositions, composed of 1 bar of background N\textsubscript{2} and CO\textsubscript{2} partial pressure ranging from 10\textsuperscript{-4} to 1 bar logarithmically, both with and without an additional modern Earth-like CH\textsubscript{4} partial pressure of $1.7$ $\mu$bar. In keeping with previous literature \citep{wolf_assessing_2017, turbet_modeling_2018, fauchez_trappist-1_2021, may_water_2021}, we do not vary the N$_2$ partial pressure and treat it as a background gas. However, we discuss the effect of different N$_2$ partial pressures on the results in section \ref{sec:limitations}.
The surface is assumed to be an aquaplanet with a slab ocean of uniform 50 meter depth, with the abundance of H\textsubscript{2}O determined by the saturation vapor pressure. The model includes sea ice and its resultant higher albedo, using the thermodynamic sea ice scheme from \citet{bitz_climate_2012}, which is standard for \texttt{ExoCAM} models. However, our sea ice scheme neglects ice drift, which has been demonstrated to affect climate and lower both surface temperature and liquid surface ocean area \citep{yang_transition_2020, yue_effect_2020}.

We use the derived planetary parameters of TRAPPIST-1e of $R_p = 5,988$ km and $g_p = 9.12$ ms$^{-2}$ from \cite{gillon_seven_2017}, and the updated stellar parameters of TRAPPIST-1 of $R_\star = 0.124 R_\odot$ and $T = 2516$ K from \cite{kane_impact_2018}. We assume TRAPPIST-1e to be spin-synchronized with an orbital and rotation period of 6.099 days, and further assume zero obliquity and zero eccentricity. 
As in \cite{may_water_2021}, we use a model stellar spectrum 
from \cite{allard_kh2_2007}, for an M dwarf with an effective temperature of 2600 K. The GCM was run until a steady state was reached in both net flux (incident stellar and thermal outwelling) and surface temperature, generally after 45$-$55 simulated Earth years.

\subsection{Spectrum Post-Processing}\label{sec:PSG}
We post-process our ExoCAM GCMs with the NASA Goddard Planetary Spectrum Generator (PSG) Global Emisson Spectra (GlobES) API\footnote{\url{https://psg.gsfc.nasa.gov/apps/globes.php}} \citep{villanueva_planetary_2018, villanueva_fundamentals_2022}, which reads in converted GCM files and uses a radiative transfer model to calculate transmission through the planetary atmosphere in each atmospheric layer. 
We then calculate a latitudinally-averaged transmission spectrum along the entire limb of the planet. 
The 
transmission spectrum is simulated using the correlated-k \verb|PUMAS| model, with 20 bins, at the full limb at longitudes of $\pm90^{\circ}$ from the substellar point. We produce 360 transit spectra taken as averages of 1 Earth day intervals 
from the last year of the GCM's temporal evolution for each atmospheric case, to investigate any effect of year-long temporal variability within the climate on variability in transmission spectra.

The spectrum is generated with an instrumental configuration analogous to that of JWST NIRSpec PRISM, with a wavelength range of 0.6$-$5.3 $\mu$m. We provide PSG with the same planetary and stellar parameters as the GCM (listed in \ref{sec:model}). In this model, limb-darkening is not considered. Atmospheric refraction is natively included, but is likely not important for TRAPPIST-1e due to the large angular size of TRAPPIST-1 \citep{Doshi:2022uu}. The full 360 transit spectra for each of the ten cases are shown in Figure \ref{fig:spectratotal}, along with the time-averaged continuum floor of each case.

To calculate the signal-to-noise ratio (SNR) of individual species, we bin our simulated PSG spectra to a resolution of $R=100$, and simulate noise for the PSG transit spectra using \verb|PandExo|\footnote{\url{https://natashabatalha.github.io/PandExo/}} \citep{batalha_pandexo_2017} for noise calculation with NIRSpec PRISM's resolution limits, as in \cite{may_water_2021}.
We vary the noise floor of NIRSpec across 0, 10, 30, and 50 ppm (as shown in Figure \ref{fig:spectratotal}), for a final set of 4 transmission spectra with varying noise floor (which supercedes the photon noise only when the photon noise is lower) from each model day in each of our 10 cases with varying pCO$_2$ and CH$_4$.

\subsection{Spectral Analysis}\label{sec:analysis}

For each atmospheric case, we individually analyze all 360 spectra to identify detectability, and its dependence on atmospheric variability. We also post-process identical GCM cases to each atmospheric case, but where molecular species are individually removed from the PSG atmospheric parameters, to produce an identical spectrum without the effects of one species. The signal-to-noise ratio of a single day for a given species is then calculated as:

\begin{equation}\label{eq:SNR}
\langle SNR \rangle = \sqrt{\sum_{\lambda} \left(\frac{m_1(\lambda) - m_2(\lambda)}{\sigma_1(\lambda)} \right)^2}
\end{equation}

Here $m_1$ is the signal strength for a single wavelength in the species-inclusive case, $m_2$ is the signal strength for the missing-species case (or the continuum), $\sigma_1$ is the noise from \verb|PandExo| at the given wavelength in the all-inclusive case, and $\sum_{\lambda}$ represents the sum over the wavelength range of the instrument \citep{lustig-yaeger_detectability_2019}. To determine the minimum number of transits for detection, 60 PSG transmission spectra separated in time by 6 days are run through \verb|PandExo|,in order to sample the time-varying climate of TRAPPIST-1e from orbit to orbit. Each spectrum is individually simulated and all 60 spectra are averaged to a mean yearly SNR, with a sequentially increasing number of transits until the mean yearly SNR surpasses $5.0$, which we consider a strong detection (99.9\% confidence), or a 50 transit limit is reached, in which case a strong detection may not be feasible.

\begin{figure}
    \centering
    \includegraphics[width=85mm]{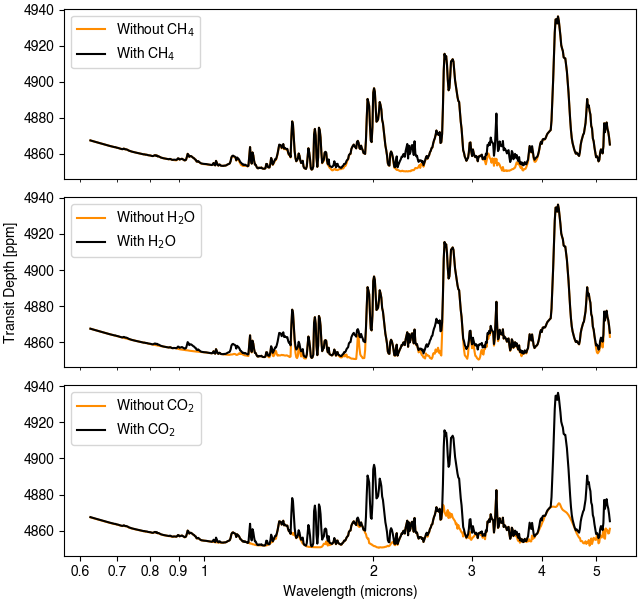}
    \caption{Spectral decomposition for the 32\textsuperscript{nd} day in the final year of the 10\textsuperscript{-2} bar pCO\textsubscript{2} CH\textsubscript{4}-inclusive case. The total transit spectrum with all species included is shown in black, with the decomposed 
    spectrum without each 
    species shown in orange. The three main peaks of CO$_2$ are visible, with the CH$_4$ features at 2.6 and 3.3 $\mu$m and the several water peaks also identifiable.}
    \label{fig:spectraspecies}
\end{figure}

NIRSpec PRISM's broad wavelength range of 0.6$-$5.3 $\mu$m \citep{jakobsen_near-infrared_2022} contains several notable spectral features of CO\textsubscript{2}, CH\textsubscript{4}, and H\textsubscript{2}O. Each species has a distinct footprint within TRAPPIST-1e's spectrum, as seen in Figure \ref{fig:spectraspecies}. 


Note that retrieval models would provide a 
more precise way of constraining atmospheric 
properties of TRAPPIST-1e from transmission spectra \citep{krissansen-totton_detectability_2018,may_water_2021,mikal-evans_detecting_2021}. However, here we simplify our analyses in order to study the daily time-resolved output from the suite of ten GCM simulations conducted in order to assess the impact of climate on observable properties of TRAPPIST-1e. 

\begin{figure*}[t]
    \centering
    \hspace*{-0.3cm}
    \includegraphics[width=185mm]{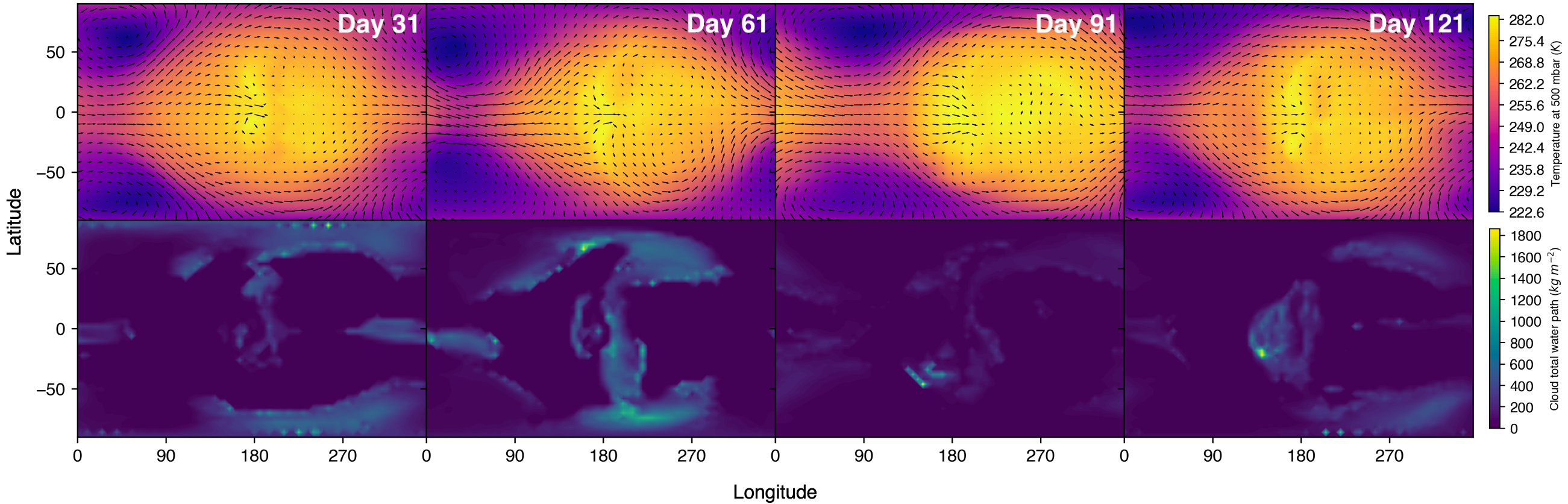}
    \caption{Four days of ExoCAM GCM simulations of TRAPPIST-1e with a partial pressure of carbon dioxide of 0.01 bar, with temperature (top) and cloud water path (bottom) at a pressure of 780 mbar. 
    There is significant cloud variability, and eastward motion of clouds due to equatorial winds which drive a hot spot offset, along with eddies on the  mid-latitude nightside.
    The substellar point (shown at the center of each plot, [0\textsuperscript{$\circ$}N, 180\textsuperscript{$\circ$}E]) is consistently cloudy and hot, but still undergoes variability in the local cloud water path.}
    \label{fig:tempcloud}
\end{figure*}

\section{Results} \label{sec:results}

\subsection{The 10\textsuperscript{-2} bar CO\textsubscript{2} Case}\label{sec:1e-2}

The 10\textsuperscript{-2} bar pCO\textsubscript{2} case has a temperate climate and we consider it as a baseline
model for the atmosphere of TRAPPIST-1e. In this section, we do a deeper dive into the analysis of this case, while noting general trends between different pCO\textsubscript{2} cases in Section \ref{sec:pco2}.

\subsubsection{Detectability of Species with JWST}\label{sec:detect}
For the methane-inclusive case, we find that only one to two transits are needed to significantly detect CO$_2$, while both H$_2$O and CH$_4$ are detectable in less than 13 transits for a noise floor of 0 or 10 ppm, although neither are detectable at noise floors of 30 ppm or higher (Table \ref{table:SNRs}). These are optimistic assumptions, as we assume a hazeless atmosphere, although, in an oxygen-free atmosphere, methane is expected to build up to higher partial pressures without oxidation into CO$_2$. However, note that we have conducted an equivalent band-by-band analysis to \cite{fauchez_impact_2019} and found good agreement, implying that haze assumptions do not greatly impact predicted feature SNR. 

\begin{table*}[]
\centering
\hspace*{-4cm}
\begin{tabular}{|cc|cc|cc|cc|cc|cc|}
\hline
\multicolumn{1}{|c|}{\multirow{2}{*}{\textbf{Species}}} & \textbf{Noise} & \multicolumn{2}{c|}{\textbf{10$^{0}$ bar pCO$_2$}}        & \multicolumn{2}{c|}{\textbf{10$^{-1}$ bar pCO$_2$}}       & \multicolumn{2}{c|}{\textbf{10$^{-2}$ bar pCO$_2$}}       & \multicolumn{2}{c|}{\textbf{10$^{-3}$ bar pCO$_2$}}       & \multicolumn{2}{c|}{\textbf{10$^{-4}$ bar pCO$_2$}}       \\ \cline{3-12} 
\multicolumn{1}{|c|}{}                                  & \textbf{Floor} & \multicolumn{1}{c|}{\textbf{CH$_4$}} & \textbf{No CH$_4$} & \multicolumn{1}{c|}{\textbf{CH$_4$}} & \textbf{No CH$_4$} & \multicolumn{1}{c|}{\textbf{CH$_4$}} & \textbf{No CH$_4$} & \multicolumn{1}{c|}{\textbf{CH$_4$}} & \textbf{No CH$_4$} & \multicolumn{1}{c|}{\textbf{CH$_4$}} & \textbf{No CH$_4$} \\ \hline
\multicolumn{1}{|c|}{\multirow{4}{*}{\textbf{CO$_2$}}}  & 0 ppm          & \multicolumn{1}{c|}{1}               & 1                  & \multicolumn{1}{c|}{1}               & 1                  & \multicolumn{1}{c|}{1}               & 1                  & \multicolumn{1}{c|}{2}               & 2                  & \multicolumn{1}{c|}{4}               & 4                  \\
\multicolumn{1}{|c|}{}                                  & 10 ppm         & \multicolumn{1}{c|}{1}               & 1                  & \multicolumn{1}{c|}{1}               & 1                  & \multicolumn{1}{c|}{1}               & 1                  & \multicolumn{1}{c|}{2}               & 2                  & \multicolumn{1}{c|}{4}               & 4                  \\
\multicolumn{1}{|c|}{}                                  & 30 ppm         & \multicolumn{1}{c|}{1}               & 1                  & \multicolumn{1}{c|}{1}               & 1                  & \multicolumn{1}{c|}{1}               & 1                  & \multicolumn{1}{c|}{2}               & 2                  & \multicolumn{1}{c|}{--}              & --                 \\
\multicolumn{1}{|c|}{}                                  & 50 ppm         & \multicolumn{1}{c|}{1}               & 1                  & \multicolumn{1}{c|}{1}               & 1                  & \multicolumn{1}{c|}{2}               & --                 & \multicolumn{1}{c|}{--}              & --                 & \multicolumn{1}{c|}{--}              & --                 \\ \hline
\multicolumn{1}{|c|}{\multirow{4}{*}{\textbf{H$_2$O}}}  & 0 ppm          & \multicolumn{1}{c|}{3}               & 4                  & \multicolumn{1}{c|}{7}               & 7                  & \multicolumn{1}{c|}{9}               & 9                  & \multicolumn{1}{c|}{14}              & 11                 & \multicolumn{1}{c|}{14}              & 12                 \\
\multicolumn{1}{|c|}{}                                  & 10 ppm         & \multicolumn{1}{c|}{5}               & 4                  & \multicolumn{1}{c|}{7}               & 7                  & \multicolumn{1}{c|}{13}              & 12                 & \multicolumn{1}{c|}{--}              & 21                 & \multicolumn{1}{c|}{--}              & 35                 \\
\multicolumn{1}{|c|}{}                                  & 30 ppm         & \multicolumn{1}{c|}{--}              & --                 & \multicolumn{1}{c|}{--}              & --                 & \multicolumn{1}{c|}{--}              & --                 & \multicolumn{1}{c|}{--}              & --                 & \multicolumn{1}{c|}{--}              & --                 \\
\multicolumn{1}{|c|}{}                                  & 50 ppm         & \multicolumn{1}{c|}{--}              & --                 & \multicolumn{1}{c|}{--}              & --                 & \multicolumn{1}{c|}{--}              & --                 & \multicolumn{1}{c|}{--}              & --                 & \multicolumn{1}{c|}{--}              & --                 \\ \hline
\multicolumn{1}{|c|}{\multirow{4}{*}{\textbf{CH$_4$}}}           & 0 ppm          & \multicolumn{1}{c|}{28}              & N/A                & \multicolumn{1}{c|}{11}              & N/A                & \multicolumn{1}{c|}{9}               & N/A                & \multicolumn{1}{c|}{9}               & N/A                & \multicolumn{1}{c|}{9}               & N/A                \\
\multicolumn{1}{|c|}{}                                  & 10 ppm         & \multicolumn{1}{c|}{--}              & N/A                & \multicolumn{1}{c|}{13}              & N/A                & \multicolumn{1}{c|}{10}              & N/A                & \multicolumn{1}{c|}{10}              & N/A                & \multicolumn{1}{c|}{9}               & N/A                \\
\multicolumn{1}{|c|}{}                                  & 30 ppm         & \multicolumn{1}{c|}{--}              & N/A                & \multicolumn{1}{c|}{--}              & N/A                & \multicolumn{1}{c|}{--}              & N/A                & \multicolumn{1}{c|}{--}              & N/A                & \multicolumn{1}{c|}{--}              & N/A                \\
\multicolumn{1}{|c|}{}                                  & 50 ppm         & \multicolumn{1}{c|}{--}              & N/A                & \multicolumn{1}{c|}{--}              & N/A                & \multicolumn{1}{c|}{--}              & N/A                & \multicolumn{1}{c|}{--}              & N/A                & \multicolumn{1}{c|}{--}              & N/A                \\ \hline
\multicolumn{2}{|c|}{\textbf{Max $\mathbf{T_{surf}}$ (K)}}                       & \multicolumn{1}{c|}{337.67}          & 338.67             & \multicolumn{1}{c|}{302.59}          & 301.64             & \multicolumn{1}{c|}{295.67}          & 295.69             & \multicolumn{1}{c|}{293.96}          & 293.61             & \multicolumn{1}{c|}{292.25}          & 290.10             \\ \hline
\multicolumn{2}{|c|}{\textbf{Min $\mathbf{T_{surf}}$ (K)}}                       & \multicolumn{1}{c|}{318.67}          & 318.33             & \multicolumn{1}{c|}{226.09}          & 214.04             & \multicolumn{1}{c|}{210.93}          & 211.07             & \multicolumn{1}{c|}{203.23}          & 193.87             & \multicolumn{1}{c|}{191.57}          & 184.25             \\ \hline
\end{tabular}
\caption{A table showing the number of transits need to achieve an SNR $\ge5.0$ for each species in each pCO\textsubscript{2} case, averaged over 6 day sample periods for 360 days simulated. A ``--" result represents cases where the SNR asymptoted to a value below $5.0$ or required more than 50 transits. The bottom two rows indicate a single Earth month time-averaged maximum and minimum surface temperature of each model.}
\label{table:SNRs}
\end{table*}

Similarly, for the zero-methane case, we find that only one transit is needed to detect CO$_2$, although it falls just short of the SNR $=5$ threshold for a noise floor of 50 ppm, with an SNR asymptoting to $4.89$. H$_2$O is detectable in 9-12 transits for a noise floor of 10 ppm or less, but undetectable for noise floor of 30 ppm or higher.

When calculating transits needed for the SNR threshold, we include the random scatter from \verb|PandExo| to model instrumental noise. However, when the additional instrumental noise is ignored and atmospheric uncertainty due to climate variability is isolated (i.e. only the photon noise from the PSG calculations is considered), 
the uncertainties within single-transit SNRs for each species ($\delta$) can be calculated as the standard deviation of the single-transit SNR across the entire year. These uncertainties are small enough that no molecule covers a range of more than $\pm1$ transits for detection between $\langle SNR \rangle_{avg} - \delta$ and $\langle SNR \rangle_{avg} + \delta$, implying that temporal variability in the $10^{-2}$ bar CO$_2$ atmosphere does not affect the detectability of species with JWST transmission spectra. This agrees with \cite{may_water_2021}, which found that temporal variability was not reflected in the detectability of species within their retrieval models with simulated NIRSpec PRISM data.  


\subsubsection{Climatological Effects on Spectra}\label{sec:climatology}

In \verb|ExoCAM|, TRAPPIST-1e has many of the expected characteristics of a Rhines rotating tidally locked planet \citep{haqq-misra_demarcating_2018}; there is a constant cloudy hotspot near the substellar point, with an eastward equatorial superrotating jet. 
There is significant variability in both the
eddies and jet, as well as cloud cover, as seen in Figure \ref{fig:tempcloud} for the $10^{-2}$ bar pCO$_2$ case. The water clouds follow a general pole-ward and eastward migration as parcels of air cool adiabatically during convection on the dayside and form clouds that are then transported by the jet to the eastern limb. 
This makes the transit spectrum of the eastern limb (270$^\circ$) much cloudier and with more muted features than the western limb (90$^\circ$). While this doesn't present an issue for JWST observations, which 
do not likely have the sensitivity to detect limb differences for terrestrial planets \citep{song_asymmetry_2021}, it is worth noting that observations of the western limb alone are more likely to have strong features of CO\textsubscript{2} and CH\textsubscript{4}.

\subsection{Dependence 
on CO\textsubscript{2} Partial Pressure}\label{sec:pco2}

An increase in the pCO\textsubscript{2} leads to an increased greenhouse effect which 
warms the deep atmosphere and surface, so higher pCO\textsubscript{2} cases are hotter and cloudier (note the higher continuum level for the $10^0$ bar case in Figure \ref{fig:spectratotal}).
This is reflected in the number of transits needed to detect H$_2$O and CH$_4$ (Table \ref{table:SNRs}). H$_2$O is detectable in fewer transits as higher CO$_2$ warms the deep atmosphere, leading to an increase in water vapor. However, CH$_4$ is muted by the cloudier atmospheres and stronger features from CO$_2$  and H$_2$O, and as such requires more transits for detection in higher pCO$_2$ cases. 

\begin{figure*}[]
    \centering
    \hspace*{-.25cm}
    \includegraphics[width=180mm]{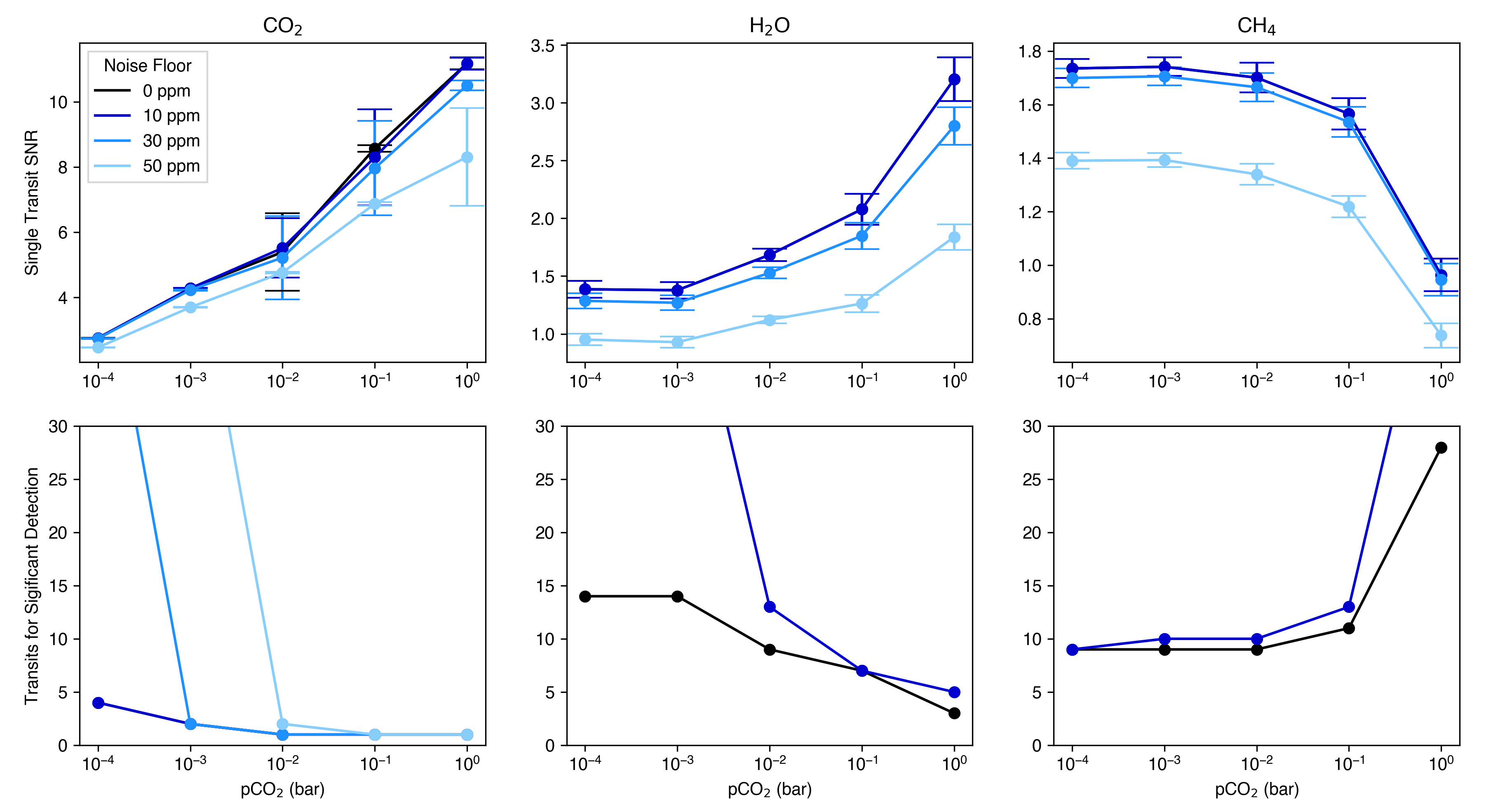}
    \caption{A cross-case comparison of the single-transit SNR (top) of each species in the atmosphere, and the resulting number of transits needed for a $5\sigma$ detection (bottom), for the methane-inclusive cases. The $\langle SNR \rangle_1$ for CO\textsubscript{2} steadily increases with pCO\textsubscript{2}, while water increases at a lower rate due to the competition between a higher self-determined mixing ratio, which increases feature height, and increased cloud cover, which flattens the spectral features. $\langle SNR \rangle_1$ for CH\textsubscript{4} decreases as its features are suppressed by the stronger CO\textsubscript{2} features, due to its decreased mixing ratio in higher pCO\textsubscript{2} cases. Note that in many cases, the 0 ppm and 10 ppm noise floor results overlap, and that in some cases, a $5\sigma$ detection is unattainable (such as in the 30 and 50 ppm CH$_4$ and H$_2$O cases), in which case the value for transits are shown beyond the axes to represent a need for over 30 transits. The full list of transits needed for detection can be found in Table \ref{table:SNRs}.}
    \label{fig:snrtransits}
\end{figure*}

While in all five cases CO\textsubscript{2} is detectable in as little as four transits for the optimistic noise floor of 0 ppm, water and methane each vary (Figure \ref{fig:snrtransits}). The number of transits needed for a $5\sigma$ detection of methane increases dramatically with pCO\textsubscript{2}, from a minimum of 9 transits in the $10^{-4}$ bar case to 28 in the $10^0$ bar case. The number of transits needed for H\textsubscript{2}O tends to decrease with increasing pCO\textsubscript{2} for the methane-inclusive case, falling from 14 transits in the $10^{-4}$ case, to 3 transits in the $10^0$ case. In all cases, as with the $10^{-2}$ bar case, there are relatively small SNR uncertainties, meaning temporal variance does not strongly affect detectability; however, the uncertainties grow with increasing pCO\textsubscript{2} (Figure \ref{fig:snrtransits}), so very hot planets may have stronger temporal variability that does regularly affect the transmission spectrum (also found by \citealp{may_water_2021}). There is also a nightside ocean ice layer in all cases, with all models except the 1 bar pCO$_2$ case also exhibiting an eyeball of ice-free ocean on the dayside, as in \citet{pierrehumbert_palette_2011}.


\section{Discussion} \label{sec:discussion}

\subsection{Feasibility of Biosignature Detection with JWST's Lifespan}

As shown in Section \ref{sec:results}, both CO\textsubscript{2} and CH\textsubscript{4} are able to be detected significantly in TRAPPIST-1e's atmosphere within 28 transits for our atmospheric cases with a noise floor of 10 ppm or less, which largely agrees with a similar analysis done by \cite{lin_differentiating_2021}. Although precisely constraining abundances would likely require further transits, 
TRAPPIST-1e is visible by JWST for about 100 days per year, which equates to about $16$ transits per year. As such, within the 5.5 year mission, both methane and carbon dioxide may be detected with a $5\sigma$ confidence level, depending on the atmospheric composition of TRAPPIST-1e and performance of NIRSpec/PRISM.

\subsection{Limitations and Further Work}
\label{sec:limitations}
In order to better characterize JWST's spectroscropic capabilities, particularly in cases including methane, it is necessary to account for haze, which we neglect in this work in order to provide optimistic constraints on methane detectability. As such, further work should include a hazy case to compare to the haze-free case, as in \cite{fauchez_impact_2019,Pidhorodetska:2020to}. CH\textsubscript{4} is assumed to be uniformly mixed and not haze-forming in our work, as well as not depleted from photochemistry or equilibrium chemistry, all of which would drastically change the transmission signal strength of both CH\textsubscript{4} and other species in the atmosphere. Notably, organic haze in the upper atmosphere could limit spectral features of other species in transmission spectra, and is less likely to form in an oxygenated atmosphere, which is expected on aquaplanets orbiting M dwarfs \citep{tian_high_2014, luger_extreme_2015}. However, in an oxygen-free environment such as our model, CH$_4$ is expected to be longer-lived and at higher concentrations, and so a low Earth-like CH$_4$ partial pressure as used in this model can indirectly approximate the effects of haze on the detectability of CH$_4$.

It is also worth mentioning that, due to TRAPPIST-1e's position on the border between two different dominant rotation dynamics, there are fundamental differences between different GCM simulations in terms of atmospheric variability timescales \citep{fauchez_trappist-1_2021, turbet_trappist-1_2021} and even basic-state climate \citep{sergeev_bistability_2022}. These can lead to disagreements of up to 50\% on the number of transits needed to reach a 5$\sigma$ spectral detection \citep{fauchez_trappist-1_2021}. \cite{fauchez_trappist-1_2021} also found that \verb|ExoCAM| produced higher cloud decks that more strongly muted features compared to other GCMs, implying that our results may be more pessimistic than those calculated with a different GCM. The model did not account for the possibility of continents, which could affect limb transmission at certain latitudes and change the water vapor abundance and cloud deck height
\citep{Lewis:2018aa,Salazar:2020aa}. It also did not include any oxygen/ozone species within the atmosphere, both of which have spectroscopic features that may overlap with our recognized species in this work \citep{lustig-yaeger_detectability_2019}.

Similarly, to truly characterize the CO\textsubscript{2}-CH\textsubscript{4} biosignature pair as biotic, as mentioned in Section \ref{sec:intro}, the existence of abundant CO 
must be ruled out. CO has two notable absorption bands in the NIRSpec PRISM range, at 4.67 $\mu$m and 2.34 $\mu$m \citep{k_n_liou_introduction_2002}, making it, or its absence, characterizable alongside CO\textsubscript{2} and CH\textsubscript{4}; the latter line would overlap with the 2.3 $\mu$m methane band, making the 4.67 $\mu$m feature imperative to carbon monoxide characterization. The 
lack of a retrieval framework precludes our ability to characterize abundance constraints and resulting surface fluxes, which is imperative to identification of any disequilibrium in the CO\textsubscript{2}$-$CH\textsubscript{4} pair when looking for biosignatures. As such, future work should apply retrieval models to the simulated transmission spectra, similarly to \cite{may_water_2021} and \cite{lin_differentiating_2021}, to better constrain the abundances and identify whether the pair exists in disequilibrium.

In order to facilitate comparison with previous work \citep{wolf_assessing_2017, turbet_modeling_2018, fauchez_trappist-1_2021}, we assume a fixed 1 bar of background N$_2$. Although we do not vary the partial pressure of N$_2$ from 1 bar in this model, we expect that an increase in N$_2$ would result in broadening of the spectral line shapes of CO$_2$, H$_2$O, and CH$_4$ due to the overall increase in atmospheric pressure, as well as more prominent collision-induced absorption (CIA) lines (and likewise, that a reduction of the N$_2$ partial pressure would decrease both). Similarly, an increase in the total surface pressure would likely result in a hotter global average surface temperature \citep{charnay_exploring_2013, wolf_controls_2014, chemke_thermodynamic_2016}. Furthermore, TRAPPIST-1e is not expected to maintain more than $\sim$1 bar of N$_2$ if it is the only atmospheric species; it would also require some amount of CO$_2$ to maintain that atmosphere against stellar wind \citep{turbet_review_2020}.
\section{Conclusions} \label{sec:conclusion}

In this work, we explored the ability of JWST to detect possible biomarkers in the atmosphere of TRAPPIST-1e, a terrestrial planet within its star's habitable zone. To do so, we simulated a grid of model atmospheres of TRAPPIST-1e with the ExoCAM GCM, consisting of N\textsubscript{2}, CO\textsubscript{2}, H\textsubscript{2}O, and CH\textsubscript{4}. We then analyzed the detectability of the CO\textsubscript{2}$-$CH\textsubscript{4} pair by simulating transmission spectra of our models with the NASA Planetary Spectrum Generator, and calculating each species' signal-to-noise ratio with \verb|PandExo|.

We find that, with optimistic limited noise (see Section \ref{sec:discussion}), CO\textsubscript{2}, H\textsubscript{2}O, and CH\textsubscript{4} would all be detectable in TRAPPIST-1e's atmosphere in as little as 13 transits for a $1$ bar N\textsubscript{2}, $10^{-2}$ bar pCO\textsubscript{2} atmosphere, with transits needed to detect CH\textsubscript{4} increasing and transits needed for H\textsubscript{2}O and CO\textsubscript{2} decreasing with increaing pCO\textsubscript{2}. Temporal variability for these atmospheres exists, but was found not to strongly affect transmission spectra or the number of transits needed for a strong detection of any molecule. Follow-up work should focus on determining the impact of limiting factors such as atmospheric haze and astronomical noise on JWST transmission spectra. 


The authors would like to thank Raymond Pierrehumbert for comments that considerably improved the manuscript. The authors also thank Eliza Kempton for helpful comments on an early draft of this manuscript, as well as Arjun Savel for help with the analysis. The authors also acknowledge the University of Maryland supercomputing resources (\url{http://hpcc.umd.edu}) and the NASA/GSFC Planetary Spectrum Generator resources (\url{https://psg.gsfc.nasa.gov/}) made available for conducting the research reported in this paper. This work was completed with resources provided by the University of Chicago Research Computing Center. 


\bibliography{ThesisSources8,References_terrestrial}{}

\end{document}